\documentclass[sii]{ipart}

\usepackage{caption}
\usepackage{subcaption}
\usepackage{amsmath}
\usepackage{graphicx,psfrag,epsf}
\usepackage{enumerate}
\usepackage{url} 
\usepackage{bm}
\usepackage{amssymb}
\usepackage{comment}
\usepackage{multirow}
\usepackage{amsmath}
\usepackage{graphicx,psfrag,epsf}
\usepackage{enumerate}
\usepackage{url} 
\usepackage{bm}
\usepackage{amssymb}
\newcommand{\code}[1]{\texttt{#1}}
\usepackage{mathtools}

\DeclarePairedDelimiter\floor{\lfloor}{\rfloor}
\usepackage{blindtext}
\usepackage{multicol}
\usepackage{color}
\usepackage{lipsum}
\usepackage{graphicx}

\RequirePackage{hyperref}

\startlocaldefs
\theoremstyle{plain}
\newtheorem{lemma}{Lemma}[section]
\newtheorem{result}{Result}[section]

\endlocaldefs

\usepackage{xcolor}
\pubyear{2021}
\volume{NA}
\issue{NA}
\firstpage{1}
\lastpage{4}
\arxiv{0000.0000}

\begin{document}

\begin{frontmatter}
\title[Correlated Wishart Matrices Classification]{Correlated Wishart Matrices Classification via an Expectation-Maximization Composite Likelihood-Based Algorithm}

\begin{aug}
    \author{\inits{Z.}\fnms{Zhou} \snm{Lan}\thanksref{t2}\ead[label=e1]{zlan@bwh.harvard.edu}}
    \address{Brigham and Women's Hospital\\
       Harvard Medical School\\
       Boston, MA 06510, USA\\
             \printead{e1}}
    \thankstext{t2}{Corresponding author.}
\end{aug}
\received{\sday{15} \smonth{8} \syear{2021}}

\begin{abstract}
Positive-definite matrix-variate data is becoming popular in computer vision. The computer vision data descriptors in the form of Region Covariance Descriptors (RCD) are positive definite matrices, which extract the key features of the images. The RCDs are extensively used in image set classification. Some classification methods treating RCDs as Wishart distributed random matrices are being proposed. However, the majority of the current methods preclude the potential correlation among the RCDs caused by the so-called auxiliary information (e.g., subjects' ages and nose widths, etc). Modeling correlated Wishart matrices is difficult since the joint density function of correlated Wishart matrices is difficult to be obtained. In this paper, we propose an Expectation-Maximization composite likelihood-based algorithm of Wishart matrices to tackle this issue. Given the numerical studies based on the synthetic data and the real data (Chicago face data-set), our proposed algorithm performs better than the alternative methods which do not consider the correlation caused by the so-called auxiliary information.
\end{abstract}

\begin{keyword}[class=AMS]
\kwd[Primary]{62H10}
\kwd[; Secondary ]{62H30}
\end{keyword}

\begin{keyword}
\kwd{Correlated Wishart Matrices}
\kwd{Composite Likelihood}
\kwd{Computer Vision}
\kwd{Expectation–maximization algorithm}
\kwd{Image Set Classification}
\kwd{Region Covariance Descriptor}
\end{keyword}

\end{frontmatter}

\section{Introduction}
\label{sec:intro}
Positive definite matrix-variate data is a type of data in which each variable is a positive definite matrix other than a scalar or a vector. The type of data is involved in several applications, e.g.,  computer vision \citep[see][]{tuzel2006region}. The computer vision data descriptor in the form of Region Covariance Descriptors (RCD) are positive definite matrices \citep{tuzel2006region}. Generally, the specification of the RCD can be given as follows. Let $I$ be an RGB color image. $I$ is usually represented as a $W \times H \times 3$ array, where $W$ and $H$ are the width and height of an image, respectively, and $3$ is the number of color channels, i.e., red (R), green (G), and blue (B). Let $F$ be the $W \times H \times D$ dimensional feature image extracted from $I$. That is, the (x,y,d) element of the array $F$ is $F(x,y,d)=\phi_d(x,y,{I})$. The function $\phi_d()$ is a mapping function to extract intensity, color, gradient, filter response, etc. at the voxel $(x,y)$ of the image $I$. Let $\bm{F}(x,y)$ be $\bm{F}(x,y)=[F(x,y,1), ..., F(x,y,D)]^T$. The RCD of the image $I$ is\footnote{In several studies, the covariance matrix is calculated within certain areas of an image.} 
\begin{equation*}
    \bm{C}_{I}=\frac{1}{WH-1}\sum_{x=1}^W\sum_{y=1}^H(\bm{F}(x,y)-\bm{\mu})(\bm{F}(x,y)-\bm{\mu})^T,
\end{equation*}
where $\bm{\mu}$ is the sample mean of all $\bm{F}(x,y)$. $\bm{C}_{I}$, a $D\times D$ covariance matrix, is a positive definite matrix-based descriptor of the image $I$. 

Compared to the traditional vector-based descriptors, many computer vision studies have found that the RCD is a very useful quantity to describe the distinguishing features of an image \citep[e.g.,][]{cherian2016bayesian}, especially for the applications to image set classification. This includes identifying objects \citep[e.g.,][]{chen2020covariance}, identifying textures \citep[e.g.,][]{diaz2009texture}, and facial recognition \citep[e.g.,][]{pang2008gabor}. Several classification methods are proposed by using the RCDs of images as inputs. {\citet{huang2015log} uses the Log-Euclidean metric to characterizing the similarities among RCDs and thus the image sets can be classified based on the Log-Euclidean distances. The most commonly used probability distribution of positive definite matrix-variate data is the Wishart distribution \citep{dryden2009non,cherian2016bayesian,lee2017inference,lan2019spatial}. Therefore, many prevailing model-based methods are proposed. For example, \citet{hidot2010expectation} proposed a Wishart mixture model relying on Expectation-Maximization algorithm. As an extension of the Wishart mixture model, the Wishart Bayesian nonparametrics method proposed by \citet{cherian2016bayesian} provides a more flexible approach for image-set classification.}

\begin{figure}[t]
    \centering
    \includegraphics[width=0.5\textwidth]{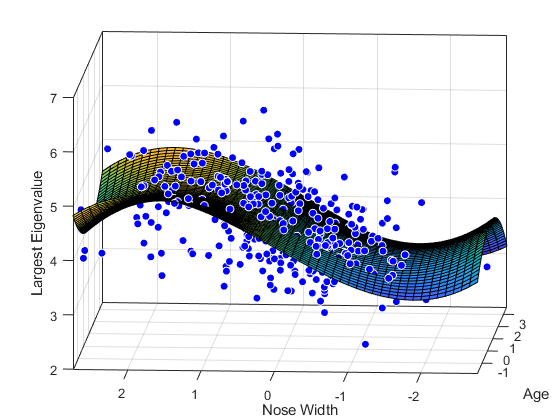}
    \caption{The 3-D scatter plot shows the RCDs' largest eigenvalues varies dependent on age and nose width. The Z-axis is for the largest eigenvalues; The X-axis and Y-axis are for the normalized ages and normalized nose widths, respectively. The points represent the observations. The fitted surface is fitted by a polynomial regression model.}
    \label{fig:movatating}
\end{figure}

Although the current works enjoy good classification, they may preclude the available auxiliary information. The so-called auxiliary information is defined as the information which is image-specific but not voxel-specific. Taking the Chicago face data-set \citep{ma2015chicago} as an example, besides the RGB images of the subjects' headshots, we also have rich image-level information such as the subjects' ages, nose widths, etc. We find that the RCDs are correlated dependent on these pieces of auxiliary information. For example, in Figure \ref{fig:movatating}, we observe the RCDs' largest eigenvalues varies dependent on age and nose width. Therefore, the RCDs are correlated across the auxiliary information. Although the correlation may or may not be explained scientifically, we conjecture that modeling the variations of the RCDs across the auxiliary information may increase the model fitness. In light of the scientific objective, the better-fitted model may also produce more accurate classification results. 

{Therefore, a model which induces the correlation among Wishart matrices with respect to the auxiliary information is preferred.} Modeling correlated Wishart matrices is difficult. The most challenging component is, unlike Gaussian distributed variables which you can feasibly use a multivariate normal distribution, the joint density function of correlated Wishart matrices is challenging to be obtained \citep{viraswami1991multivariate,blumenson1963properties,smith2007distribution}. A desirable but challenging {approach} is the so-called (spatial) Wishart process \citep{gelfand2004nonstationary,lan2021geostatistical}. The model has a very elegant construction, that is to induce dependence of Wishart matrices via the latent Gaussian processes. Unfortunately, this density function of the (spatial) Wishart process is not analytically available under general conditions \citep{viraswami1991multivariate,blumenson1963properties,smith2007distribution}. Fortunately, standing on the shoulders of the giants who provided the prodigious results \citep{james1964distributions}, we can obtain the analytic expression of the \textit{bivariate} density function.

To mitigate this inference issue caused by the analytic form of the joint density function, we propose a computationally-feasible composite likelihood-based inference equipped with an expectation-maximization (EM) algorithm. We first propose a hierarchical model based on the (spatial) Wishart process. Given our proposed hierarchical model, we use the EM algorithm for model fitting. However, due to the analytic form of the joint density function is not available, both the E-step and M-step cannot be executed in a standard way. Therefore, we replace the full likelihood with the composite likelihood in the algorithm. The composite likelihood is an inference function derived by multiplying a collection of component likelihood \citep{lindsay1988composite}, which is primarily to resolve the issue that the full likelihood is unavailable. Because the analytic density function of two correlated Wishart matrices can be obtained \citep{james1964distributions}, the composite likelihood is in a pairwise setting, which is then derived by multiplying a collection of possible pairs. The estimators based on composite likelihood is asymptotically unbiased \citep{varin2011overview}, but the efficiency of the estimators is determined by the likelihood-weights \citep{padoan2010likelihood,bevilacqua2012estimating,bai2012joint,bai2014efficient}. Considering a trade-off between parameter estimation and computational cost, we further proposed a novel likelihood-weight function, enjoying accurate parameter estimation and fast computational speed.

In light of our goal of classification, we use both the simulated data and the real data to demonstrate that our proposed method produces a more accurate classification of positive definite matrices than the benchmark methods. The simulated data are generated from the proposed model. The real data is the Chicago face data \citep{ma2015chicago}. The benchmark methods include a wide range of methods, including Wishart Bayesian nonparametrics (Wishart BNP) \citep{cherian2016bayesian}, Log-Euclidean metric method \citep{huang2015log}, K-means, and Gaussian mixture model. The numerical results show that our proposed method produces a more accurate classification of positive definite matrices, demonstrating its compelling potential in image set classification in the coming future. 

In the rest of the paper, we first give our hierarchical model in Section \ref{sec:setup}. The estimation method is provided in Section \ref{sec:estimation}. We provide simulation studies in Section \ref{sec:numerical}, showing that our algorithm performs better than the other benchmark methods. A real data application using the Chicago face database \citep{ma2015chicago} with the objective of race classification is provided in Section \ref{sec:application}. In Section \ref{sec:conclusion}, we conclude with a discussion. The essential derivations and the codes written in \code{MATLAB} for implementing our method are given in the appendices.

\section{Model}
\label{sec:setup}
In this section, we illustrate our model construction step-by-step. First, we give the basic model setup in Section \ref{sec:base}. The basic model describes how the correlation of positive definite matrices across the auxiliary information is constructed. In Section \ref{sec:latent}, we further introduce the latent variables which can be used for the purpose of classification. Finally in Section \ref{sec:final}, we give a model summary and introduce its relationship to the RCD-based image set classification.

Our model is based on the the parameterized Wishart distribution. Note that the \textit{parameterized} Wishart distribution is slightly different from the Wishart distribution introduced in the classic textbooks. The parameterized Wishart distribution is constructed as follows. $\bm{G}_j$ independently follows a mean-zero normal distribution with the covariance matrix $\bm{\Sigma}$, denoted as $\bm{G}_j\stackrel{ind}{\sim}\mathcal{N}(\bm{0},\bm{\Sigma})$. $\bm{W}=\sum_{j=1}^M\bm{G}_j\bm{G}_j^T/M$ follows a parameterized Wishart distribution with the mean matrix $\bm{\Sigma}$ and the degrees of freedom $M$, denoted as $\bm{W}\sim\mathcal{W}_p(\bm{\Sigma},M)$. The density function of $\bm{W}$ is given below
\begin{equation}
\label{eq:pdf}
    {\displaystyle f_{\mathbf {W} }(\mathbf {w} )={\frac {|\mathbf {w} |^{(M-p-1)/2}e^{-\operatorname {tr} (M\mathbf {\Sigma} ^{-1}\mathbf {w} )/2}}{2^{\frac {Mp}{2}}|{\mathbf {\Sigma}/M }|^{M/2}\Gamma _{p}({\frac {M}{2}})}}}.
\end{equation}

\subsection{Basic Model Setup}
\label{sec:base}
Let $\bm{A}_t$ be a $p\times p$ positive definite matrix-random variable following a parameterized Wishart distribution with the mean matrix $\bm{\Sigma}_t$ and the degrees of freedom $M$, denoted as 
\begin{equation}
    \bm{A}_t\sim\mathcal{W}_p(\bm{\Sigma}_t,M).
\end{equation}
$\bm{A}_t$ can be treated as the RCD of the image $t$. Our goal is to build a model which induces the correlation between $\bm{A}_t$ and $\bm{A}_s$ ($t\not=s$). To achieve this goal,  We decompose $\bm{A}_t$ as $\bm{A}_t=\bm{L}_t\bm{U}_t\bm{L}_t^T$, where $\bm{L}_t$ is the lower Cholesky factor of $\bm{\Sigma}_t$ such as $\bm{\Sigma}_t=\bm{L}_t\bm{L}_t^T$. Since $\bm{\Sigma}_t$ determines the mean of $\bm{A}_t$, we treat $\bm{U}_t$ as the terms inducing the correlation. A more approachable interpretation of this decomposition is that $\bm{\Sigma}_t$ is the \textit{mean} term driving the mean and $\bm{U}_t$ is the \textit{remaining} term explaining the correlation among the matrices \citep{lan2021geostatistical}. Next, we give the details of constructing these two terms.


{A popular approach} to construct the correlation among $\bm{U}_t$ has been known as (spatial) Wishart process \citep{gelfand2004nonstationary,lan2021geostatistical}. {The \textit{standard} Wishart process is defined as follows. Let $\bm{G}_t$ be a mean-zero Gaussian process with the covariance as $\text{cov}(\bm{G}_t,\bm{G}_s)=\rho_{ts}\times\bm{I}$, where $\rho_{ts}\in(0,1)$ is the correlation coefficient and $\bm{I}$ is an identity matrix.} If we have independent and identically distributed realizations of this process, denoted as $\bm{G}_{jt}$, where $j$ is the index for realizations and $t$ is the index for the Gaussian process, then $\bm{U}_t=\sum_{j=1}^M\bm{G}_{jt}\bm{G}_{jt}^T/M$ follows a so-called Wishart process, denoted as $\{\bm{U}_t\}_{t\in\mathbb{T}}\sim\mathcal{WP}(M,\rho_{ts}{,}\bm{I})$, where $\mathbb{T}$ is a set of indices for images. The correlation $\rho_{ts}$ can be expressed as a correlation function, i.e., $\rho_{ts}=\mathcal{K}(||\bm{X}_t-\bm{X}_s||;\bm{\phi})$, where $\mathcal{K}(d;\bm{\phi})$ is a correlation function (e.g., Matern, exponential) of $d$, and $\bm{\phi}$ is a vector of the function parameters. Let $\bm{X}_t$ be a $d\times 1$ vector representing the auxiliary information of the image $t$ (e.g., age, nose width, etc). The distance $||\bm{X}_t-\bm{X}_s||$ is the Euclidean distance of the two images' auxiliary information. 

We emphasize that this model construction has two important properties: (1) $\bm{A}_t$ preserves to follow $\mathcal{W}_p(\bm{\Sigma}_t,M)$ marginally; (2) $\bm{A}_t$ and $\bm{A}_s$ are correlated ($t\not=s$) due to the \textit{remaining} terms. The property (1) can be validated by giving the lemma below.

\begin{lemma}
\label{lemma:1}
Let $\bm{\Sigma}$ be a $p\times p$ positive definite matrix. $\bm{L}$ is the lower Cholesky factor of $\bm{\Sigma}$ such as $\bm{\Sigma}=\bm{L}\bm{L}^T$. If $\bm{U}\sim \mathcal{W}_p(\bm{I},M)$, then $\bm{W}=\bm{L}\bm{U}\bm{L}^T\sim  \mathcal{W}_p(\bm{\Sigma},M)$.
\end{lemma}
This lemma is borrowed from Proposition 8.1 of \citet[][Chapter 8: The Wishart Distribution]{eaton1983multivariate} with a proof, and it can also be found in many other multivariate textbooks. To validate the property (2), we can quantify the dependence of the Wishart process by using the expected squared Frobenius norm {\citep[][Equation 3]{lan2021geostatistical}}, expressed as
\begin{equation}
    \begin{aligned}
    &\mathbb{E}||\bm{U}_t-\bm{U}_s||_F^2=\gamma(M)(1-\rho_{ts}^2)\\
    & {\gamma(M)=\frac{2}{M}\Big(p+p^2\Big)},
    \end{aligned}
\end{equation}
where $||\bm{U}_t-\bm{U}_s||_F^2$ is the squared Frobenius norm of the term $\bm{U}_t-\bm{U}_s$. The norm can also be treated as a variogram \citep{cressie1992statistics}. Given the variogram {which measures Euclidean distance of two positive definite matrices}, it is more transparent to claim that a larger correlation of the underlying Gaussian process leads to a larger dependence of {the random Wishart matrices}, and a larger value of $M$ leads to a smaller local variation {of the random Wishart matrices}.

\subsection{Latent Variable Setup}
\label{sec:latent}
In light of our scientific objective to classify the RCDs, the mean matrix $\bm{\Sigma}_t$ is specified as a term dependent on a latent group variable. It is specified as $\bm{\Sigma}_t|[Z_t=k]\equiv \mathbf{S}_k$, where $Z_t\in\{1,2,...,K\}$ is the latent group index. The latent group variable $Z_t$ independently follows a discrete distribution such that $Pr(Z_t=k)=\omega_k$, denoted as
\begin{equation}
    Z_t\stackrel{ind}{\sim}\text{Discrete}([1,2,..,K],[\omega_1,\omega_2, ..., \omega_K]).
\end{equation}

The hyperparameters $\{\mathbf{S}_k: k=1,2,...,K\}$ are the known and given values describing the features of each group ($k\in\{1,2,...,K\}$) in terms of positive definite matrices. The hyperparameters $\{\mathbf{S}_k: k=1,2,...,K\}$ can be obtained from the training data and/or the experienced values. Because the hyperparameters $\{\mathbf{S}_k: k=1,2,...,K\}$ are known and thus the images are classified in a supervised manner, we refer the process of identifying which of a set of categories an image belongs to as \textit{classification}. The number of components $K$ is also given and fixed. Based on this hierarchical model setup, $Z_t$ can be used as a group indicator to classify the image $t$ to a group $k\in\{1,2,...,K\}$.

\subsection{Model Summary and the RCD-Based Image Set Classification}
\label{sec:final}
To this end, our proposed hierarchical model framework is summarized as
\begin{equation}
\label{eq:model}
    \begin{aligned}
    &\textbf{Basic Level}:\ \\
    &\bm{A}_t=\bm{L}_t\bm{U}_t\bm{L}_t^T\quad \bm{\Sigma}_t=\bm{L}_t\bm{L}_t^T\quad\\ &\{\bm{U}_t\}_{t\in\mathbb{T}}\sim\mathcal{WP}(M,\rho_{ts}{,}\bm{I})\quad \rho_{ts}=\mathcal{K}(||\bm{X}_t-\bm{X}_s||;\bm{\phi})\\
    \\
    &\textbf{Latent Level:}\ \\
    &\bm{\Sigma}_t|[Z_t=k]\equiv \mathbf{S}_k\quad\\
    &Z_t\stackrel{ind}{\sim}\text{Discrete}([1,2,..,K],[\omega_1,\omega_2, ..., \omega_K]).
    \end{aligned}
\end{equation}

The above illustration gives a classification model for the correlated positive definite matrices, which may be still abstract to the practitioners. Next, we elaborate our proposed model by giving its relationship to the RCD-based image set classification. The image set classification is an important research field in computer vision. The objective is to classify the pictures into different topics \citep[e.g.,][]{cherian2016bayesian}. We use the Chicago face database \citep{ma2015chicago} as an illustration example. The data-set provides high-resolution, standardized headshots of male and female faces of different races between the ages of 17-65, as well as the subjects' auxiliary information (e.g., age, nose width, etc). We have a scientific goal to classify the pictures by their races, i.e.,  Black, Latino, and White, which is one of the common scientific goals in computer vision studies \citep[e.g.,][]{ou2005real,roomi2011race,fu2014learning}. 

Let $\bm{A}_t$ be the RCD of the face picture $t$. First, the \textit{remaining} terms in the basic level capture the variations caused by the auxiliary information. Because the auxiliary information (e.g., age, nose width, etc) is usually irrelevant to the races, our model specification indicates that the dependence is separated from the effect of races which are driven by the \textit{mean} terms. 

Given that the potential correlation among the random positive definite matrices caused by the auxiliary information has been captured by our proposed model, we can classify the data with fewer \textit{noises}. We use $k=\{1,2,3\}$ to represent  Black, Latino, and White, respectively. The hyperparameters $\{\mathbf{S}_k: k=1,2,...,K\}$ are plug-in values that can be obtained from the training data and/or the experienced value, describing the typical features of a race in terms of RCD. The maximum a posteriori estimator of $Z_t$ can be used for the classification of images, resolving the scientific objective.



\section{Estimation}
\label{sec:estimation}
In this section, we describe our approach to obtain the parameter estimates of the unknowns. Our estimation approach is essentially a hybrid method composed of the composite likelihood method and the EM algorithm.

\subsection{Standard EM Algorithm}
Given Model (\ref{eq:model}), we first provide the EM Algorithm under the standard situations. Rigorously, we use $[\bm{a}_1, ...,\bm{a}_t]$ to represent the observed positive definite matrices of the random matrices $[\bm{A}_1, ...,\bm{A}_t]$. We give that $\bm{a}=[\bm{a}_1, ...,\bm{a}_t]$ and $\bm{A}=[\bm{A}_1, ...,\bm{A}_t]$. We also give $\bm{Z}=[Z_1, ..., Z_T]$ to denote all latent group indexes. The parameters to be estimated are $\bm{\omega}=[\omega_1, ..., \omega_k, ..., \omega_{K}]$ and $\bm{\Phi}=[\bm{\phi}^T, M]^T$, where $[\omega_1, ..., \omega_k, ..., \omega_{K}]$ are the group-weights, $\bm{\phi}$ is a vector of the parameters in the correlation function $\mathcal{K}(d;\bm{\phi})$, and $M$ is the degrees of freedom. We give the notation $\bm{\theta}=[M, \bm{\omega}^T, \bm{\Phi}^T]^T$ which contains all the parameters needed to be estimated. The full log-likelihood is 
\begin{equation}
    \begin{aligned}
   \ell_{Full}(\bm{\theta}|\bm{a},\bm{Z})&= \log\Bigg[\prod_{j_1=1}^K ...\prod_{j_t=1}^K ... \prod_{j_T=1}^K \\
   &\bigg(f_{\bm{A}}(\bm{a}|\bm{\Sigma}_1=\bm{S}_{j_1}, ..., \bm{\Sigma}_t=\bm{S}_{j_t}, ...,\bm{\Sigma}_T=\bm{S}_{j_T}, \bm{\Phi})\\
    &\times \prod_{t=1}^T\omega_{j_t}\bigg)^{\mathbb{I}(Z_1=j, , ..., Z_t=j_t, ..., Z_T=j_T)}\Bigg],
    \end{aligned}
\end{equation}
where $f_{\bm{A}}(\bm{a}|\bm{\Sigma}_1=\bm{S}_{j_1}, ..., \bm{\Sigma}_t=\bm{S}_{j_t}, ...,\bm{\Sigma}_T=\bm{S}_{j_T}, \bm{\theta})$ is the joint density function of $\bm{A}$ conditional on that $[Z_1=j_1, ..., Z_t=j_t , ..., Z_T=j_T]$. 

Our goal is to obtain the maximum likelihood estimates for $\bm{\theta}$. Therefore, we have to maximize the likelihood whose the latent variables $\bm{Z}$ are integrated out. The log-likelihood $\ell(\bm{\theta}|\bm{a})$ whose latent variables are integrated out is expressed as
\begin{equation}
    \begin{aligned}
   \ell(\bm{\theta}|\bm{a})&= \int ... \int ...\int  \log\Bigg[\prod_{j_1=1}^K ...\prod_{j_t=1}^K ... \prod_{j_T=1}^K \\
   &\bigg(f_{\bm{A}}(\bm{a}|\bm{\Sigma}_1=\bm{S}_{j_1}, ..., \bm{\Sigma}_t=\bm{S}_{j_t}, ..., \bm{\Sigma}_T=\bm{S}_{j_T}, \bm{\Phi})\\
     &\times \prod_{t=1}^T\omega_{j_t}\bigg)^{\mathbb{I}(Z_1=j, ...,Z_t=j_t , ..., Z_T=j_T)}\Bigg]dZ_1 ... dZ_t ... dZ_T,
    \end{aligned}
\end{equation}

The hurdle caused by the unobserved latent variables $\bm{Z}$ can be easily overcome through the famous EM algorithm. The iterative scheme of the algorithm is given as follows. If we have the current parameter $\bm{\theta}^{(r)}$, the likelihood is $\ell_{Full}(\bm{\theta}^{(r)}|\bm{a},\bm{Z})$. We first compute the expected value of the likelihood function with respect to the current conditional distribution of $\bm{Z}$ given the data $\bm{a}$ and the current parameter estimates $\bm{\theta}^{(r)}$, denoted as $Q(\bm{\theta}|\bm{\theta}^{(r)})=\mathbb{E}_{\bm{Z}|\bm{A},\bm{\theta}^{(r)}} \ell_{Full}(\bm{\theta}|\bm{a},\bm{Z})$. We call this step as E-step since we take the expectation. Next we maximize the term $Q(\bm{\theta}|\bm{\theta}^{(r)})$ with respect to $\bm{\theta}$ to obtain the next estimates $\bm{\theta}^{(r+1)}$. This step is called M-step because we maximize the function. We iteratively repeat the two steps until convergence. Finally, $\hat{\bm{\theta}}$ is obtained as the maximum likelihood estimates for $\bm{\theta}$.

\subsection{Composite Likelihood-Based Inference}
The standard EM algorithm has a bottleneck which is caused by the joint density function $f_{\bm{A}}(\bm{a}|\bm{\Sigma}_1=\bm{S}_{j_1}, ..., \bm{\Sigma}_t=\bm{S}_{j_t}, ..., \bm{\Sigma}_T=\bm{S}_{j_T}, \bm{\Phi})$. Given several reports \citep{viraswami1991multivariate,blumenson1963properties,smith2007distribution}, an analytic expression of the joint density function is not available under general conditions. However, standing on the shoulders of the giants who provided the prodigious results \citep{james1964distributions}, we are able to obtain the \textit{bivariate} density, i.e., the conditional density function of $[\bm{A}_t, \bm{A}_s]$ given $[Z_t, Z_s]$, denoted as $f_{\bm{A}_t,\bm{A}_s}(\bm{a}_t,\bm{a}_s|\bm{\Sigma}_t=\bm{S}_{j_t},\bm{\Sigma}_s=\bm{S}_{j_s},\bm{\Phi})$. The details in obtaining the density function are summarized in Appendix \ref{sec:derivation}. The expression of $f_{\bm{A}_t,\bm{A}_s}(\bm{a}_t,\bm{a}_s|\bm{\Sigma}_t=\bm{S}_{j_t},\bm{\Sigma}_s=\bm{S}_{j_s},\bm{\Phi})$ is written as
{
\begin{equation}
\small
    \begin{aligned}
&f_{\bm{A}_t,\bm{A}_s}(\bm{a}_t,\bm{a}_s|\bm{\Sigma}_t=\bm{S}_{j_t},\bm{\Sigma}_s=\bm{S}_{j_s},\bm{\Phi})\\
    &=\ _0F_1\left(\frac{1}{2}M;\frac{1}{4}\left(\frac{M\rho_{ts}}{1-\rho_{ts}^2}\right)^2\bm{Q}_{j_s}^{-1}\bm{a}_s(\bm{Q}_{j_s}^T)^{-1}\bm{Q}_{j_t}^{-1}\bm{a}_t(\bm{Q}_{j_t}^T)^{-1}\right)\\
    & 2^{-Mp}  \left[\Gamma_p(\frac{1}{2}M)\right]^{-2} |\bm{S}_{j_t}^{-1}\bm{a}_t|^{\frac{1}{2}(M-p-1)} |\bm{S}_{j_s}^{-1}\bm{a}_s|^{\frac{1}{2}(M-p-1)} \\
    &\left(\frac{1-\rho_{ts}^2}{M^2}\right)^{-\frac{pM}{2}}  \exp\left(-\frac{1}{2}tr\left[M\frac{1}{1-\rho_{ts}^2}\left(\bm{S}_{j_s}^{-1}\bm{a}_s+\bm{S}_{j_t}^{-1}\bm{a}_t\right)\right]\right)\\
    & |\bm{S}_{j_t}^{-1}|^{\frac{p+1}{2}} |\bm{S}_{j_s}^{-1}|^{\frac{p+1}{2}},
    \end{aligned}
\end{equation}
}
where $_0F_1(;)$ is a hypergeometric function of matrix argument \citep[See][Section 6]{james1964distributions} and $\bm{Q}_{j_t}$ is the lower Cholesky decomposition factor of $\bm{S}_{j_t}$ such as $\bm{S}_{j_t}=\bm{Q}_{j_t}\bm{Q}_{j_t}^T$. The value of this function $_0F_1(;)$ can be numerically evaluated \citep{koev2006efficient}. 

Here, we want to highlight two important properties about this density function: (1) when $\rho_{ts}=0$, $f_{\bm{A}_t,\bm{A}_s}(\bm{a}_t,\bm{a}_s|\bm{\Sigma}_t=\bm{S}_{j_t},\bm{\Sigma}_s=\bm{S}_{j_s},\bm{\Phi})$ is the product of the probability density functions of $\mathcal{W}_p(\bm{S}_{j_t},M)$ and $\mathcal{W}_p(\bm{S}_{j_s},M)$ (see the density function of parameterized Wishart distribution in Equation \ref{eq:pdf}); (2) conversely, $\rho_{ts}=1$ leads to that $f_{\bm{A}_t,\bm{A}_s}(\bm{a}_t,\bm{a}_s|\bm{\Sigma}_t=\bm{S}_{j_t},\bm{\Sigma}_s=\bm{S}_{j_s},\bm{\Phi})$ is an improper density function. The property (1) is consistent with our basic model setup because $\rho_{ts}=0$ implies that there is no dependence. The property (2) may be analogised to the similar issue encountered in Gaussian process modeling \citep{williams2006gaussian}, that $\rho_{ts}=1$ leads to the improper multivariate Gaussian density function.

Obtaining the expression of the \textit{bivariate} joint density is a milestone which allows us to proceed the next composite likelihood-based inference. Composite likelihood is an inference function derived by multiplying a collection of component likelihood. For example, we denote ${\mathcal{A}_1, ..., \mathcal{A}_C}$ as a set of marginal or conditional events with associated likelihoods $\mathcal{L}_c(\bm{\theta};\bm{y})\propto f(\bm{y}\in \mathcal{A}_c;\bm{\theta})$, where $f()$ is a density function, $\bm{y}$ is the data, and $\bm{\theta}$ is the parameters. Following \citet{lindsay1988composite}, a composite likelihood is a weighted product such as $\prod_{c=1}^C\mathcal{L}_c(\bm{\theta};\bm{y})^{p_c}$, where $\{p_c: k=1,2,...,C\}$ are non-negative weights. Under the usual regularity conditions, the composite likelihood-based inference can provide asymptotically unbiased parameter estimates when standard likelihood estimators are not available \citep{varin2011overview}. For example, \citet{padoan2010likelihood} proposed a pairwise setting for max-stable processes by using composite likelihood-based inference, since only the analytically tractable form can be obtained in their case. 

We define the events $\mathcal{A}_c=\{\bm{A}_t,\bm{A}_s\}$ as the sets of \textit{bivariate} positive definite matrices taken over all $T(T-1)/2$ distinct pairs $t$ and $s$. Next we replace the full log-likelihood $\ell_{Full}(\bm{\theta}|\bm{a},\bm{Z})$ with the composite log-likelihood $\ell_{Pair}(\bm{\theta}|\bm{a},\bm{Z})$. The log composite likelihood $\ell_{Pair}(\bm{\theta}|\bm{a},\bm{Z})$ in the pairwise setting is written as 
\begin{equation}
    \begin{aligned}
    \ell_{Pair}(\bm{\theta}|\bm{a},\bm{Z})=\sum_{t=1}^{T-1}\sum_{s=t+1}^{T}p_{ts}\ell_{ts}(\bm{\theta}|\bm{a},\bm{Z}),
    \end{aligned}
\end{equation}
where $\ell_{ts}(\bm{\theta}|\bm{a},\bm{Z})=\log\Bigg[\prod_{j_t=1}^K \prod_{j_s=1}^K \bigg( f_{\bm{A}_t,\bm{A}_s}(\bm{a}_t,\bm{a}_s|\bm{\Sigma}_t=\bm{S}_{j_t},\bm{\Sigma}_s=\bm{S}_{j_s},\bm{\Phi})\times \omega_{j_t}\omega_{j_s}\bigg)^{\mathbb{I}(Z_t=j_t, Z_s=j_s)}\Bigg]$ and $p_{ts}>0$ is its likelihood-weight. We usually scale the likelihood-weights so that $\sum_{t<s}p_{ts}=1$.


The estimators obtained via composite likelihood-based inference are asymptotically unbiased but not asymptotically efficient \citep{varin2011overview,padoan2010likelihood}. A key to resolve this issue is the choice of the likelihood-weights $\{p_{ts}\}$ that minimize the total amount of variation of the asymptotic variance \citep[][Section 2.3]{heyde2008quasi}. {In theory, the  weights of composite likelihood inference are to increase the efficiency of parameter estimation (i.e., asymptotic relative efficiency ARE). Intuitively, it is to put more weights on the more correlated pairs is the \textit{golden rule}. There have been two prevailing methods to specify the weights: \citet{padoan2010likelihood} proposed the $0/1$ likelihood-weights where a threshold $\delta>0$ is introduced. If $||\bm{X}_t-\bm{X}_s||>\delta$, then $p_{ts}=0$, otherwise, $p_{ts}=1$ (not scaled). This method is computationally-feasible because many pairs are excluded from computation. \citet{bai2012joint} and \citet{bevilacqua2012estimating} further propose adaptive weights which rely on the analytic expression of the Fisher information of estimators. The first one provides a computationally feasible method but sacrifices efficiency. The latter method usually provides the best efficiency by giving smooth weights, however, in our case, it is non-trial to obtained since the close form of the Fisher information cannot be easily obtained due to the presence of hypergeometric function of matrix argument. }


To combine the two ideas above, we propose a novel \textit{smooth} likelihood-weight function which is simple and computationally-fast for the practitioners but produces relatively efficient estimators. The proposed likelihood-weights are $p_{ts}\propto\exp(-||\bm{X}_t-\bm{X}_s||/\lambda)\times u_{ts}$. The variable $u_{ts}\in\{0,1\}$ is pre-specified: among all $u_{ts}$, we randomly select $\floor*{u\times T(T-1)/2}$ pairs which have $u_{ts}=1$, where $u\in (0,1)$; the rest pairs have $u_{ts}=0$. This likelihood-weight function has two tuning parameters. $\lambda>0$ controls the decay rate of the likelihood-weight. That is, a smaller $\lambda$ leads to more homogeneous non-zero likelihood-weights (See Figure \ref{fig:exp}). {More importantly, for any $\lambda$, the pairs of observations that are irrelevant apart in auxiliary information are smoothly down-weighed, which is preferred as outlined in \citet{bevilacqua2012estimating} .} $u\in (0,1)$ controls the computational cost and reduces the pairs to be evaluated from $T(T-1)/2$ to $u\times T(T-1)/2$. In the numerical studies, we will investigate the roles of the two tuning parameters on parameter estimation and classification accuracy.

\subsection{Composite Likelihood-Based EM Algorithm}
Combining all the results above, we give the final EM composite likelihood-based algorithm, which is a hybrid method composed of the EM algorithm and composite likelihood-based inference. The key idea of this hybrid method is to replace $\ell_{Full}(\bm{\theta}|\bm{a},\bm{Z})$ by $\ell_{Pair}(\bm{\theta}|\bm{a},\bm{Z})$ in the algorithm. Thus, the step of conditional expectation is with the posterior based on composite likelihood, that is $p(\bm{Z}|\bm{A},\bm{\theta}^{(r)})\propto \ell_{Pair}(\bm{\theta}^{(r)}|\bm{a},\bm{Z})p(\bm{Z})$. The corresponding E-step and M-step are described as follows:
\textbf{E-Step:} In this step, we aim to obtain
$\mathbb{E}_{\bm{Z}|\bm{A},\bm{\theta}^{(r)}}\ell_{Pair}(\bm{\theta}|\bm{a},\bm{Z})$. It is 

    \begin{equation}
    \small
    \begin{aligned}
     &Q_{Pair}(\bm{\theta}|\bm{\theta}^{(r)})=\mathbb{E}_{\bm{Z}|\bm{A},\bm{\theta}^{(r)}}\ell_{Pair}(\bm{\theta}|\bm{a},\bm{Z})\\
     &=\mathbb{E}_{\bm{Z}|\bm{A},\bm{\theta}^{(r)}}\Bigg(\sum_{t=1}^{T-1}\sum_{s=t+1}^{T}p_{ts}\log\Bigg[\prod_{j_t=1}^K\prod_{j_s=1}^K\\
     &  \bigg( f_{\bm{A}_t,\bm{A}_s}(\bm{a}_t,\bm{a}_s|\bm{\Sigma}_t=\bm{S}_{j_t}, \bm{\Sigma}_s=\bm{S}_{j_s}, \bm{\Phi})\times \omega_{j_t}\omega_{j_s}\bigg)^{\mathbb{I}(Z_t=j_t, Z_s=j_s)}\Bigg]\Bigg)\\
     &=\mathbb{E}_{\bm{Z}|\bm{A},\bm{\theta}^{(r)}}\Bigg(\sum_{t=1}^{T-1}\sum_{s=t+1}^{T}p_{ts}\sum_{j_t=1}^K \sum_{j_s=1}^K\\
     &\log\bigg( f_{\bm{A}_t,\bm{A}_s}(\bm{a}_t,\bm{a}_s|\bm{\Sigma}_t=\bm{S}_{j_t}, \bm{\Sigma}_s=\bm{S}_{j_s}, \bm{\Phi})\times \omega_{j_t}\omega_{j_s}\bigg)\\
     &{\mathbb{I}(Z_t=j_t, Z_s=j_s)}\Bigg)\\
     &=\sum_{t=1}^{T-1}\sum_{s=t+1}^{T}p_{ts}\sum_{j_t=1}^K \sum_{j_s=1}^K \\
     &\log\bigg( f_{\bm{A}_t,\bm{A}_s}(\bm{a}_t,\bm{a}_s|\bm{\Sigma}_t=\bm{S}_{j_t}, \bm{\Sigma}_s=\bm{S}_{j_s}, \bm{\Phi})\times \omega_{j_t}\omega_{j_s}\bigg)\\
     &\mathbb{E}_{\bm{Z}|\bm{A},\bm{\theta}^{(r)}}{\mathbb{I}(Z_t=j_t, Z_s=j_s)}.
    \end{aligned}
\end{equation}

This means that we only need to calculate $T_{\binom{j_t,j_s}{t,s}}^{(r)}=\mathbb{E}_{\bm{Z}|\bm{A},\bm{\theta}^{(r)}}\mathbb{I}(Z_t=j_t, Z_s=j_s)$ and plug them in. The term $T_{\binom{j_t,j_s}{t,s}}^{(r)}$ is explicitly expressed as
\begin{equation}
\small
    \begin{aligned}
    & T_{\binom{j_t,j_s}{t,s}}^{(r)}=\mathbb{E}_{\bm{Z}|\bm{A},\bm{\theta}^{(r)}}\mathbb{I}(Z_t=j_t, Z_s=j_s)\\
    &=\Big(f_{\bm{A}_t,\bm{A}_s}(\bm{a}_t,\bm{a}_s|\bm{\Sigma}_t=\bm{S}_{j_t}, \bm{\Sigma}_s=\bm{S}_{j_s}, \bm{\Phi}^{(r)}) \omega_{j_t}^{(r)}\omega_{j_s}^{(r)}\Big)^{p_{ts}}\div\\
    &\sum_{j_t=1}^K \sum_{j_s=1}^K \bigg( f_{\bm{A}_t,\bm{A}_s}(\bm{a}_t,\bm{a}_s|\bm{\Sigma}_t=\bm{S}_{j_t}, \bm{\Sigma}_s=\bm{S}_{j_s}, \bm{\Phi}^{(r)}) \omega_{j_t}^{(r)}\omega_{j_s}^{(r)}\bigg)^{p_{ts}}.
    \end{aligned}
\end{equation}

    By plugging in $T_{\binom{j_t,j_s}{t,s}}^{(r)}$, we have the updated $Q_{Pair}(\bm{\theta}|\bm{\theta}^{(r)})$ expressed as 
    \begin{equation}
    \begin{aligned}
     &Q_{Pair}(\bm{\theta}|\bm{\theta}^{(r)})=\mathbb{E}_{\bm{Z}|\bm{A},\bm{\theta}^{(r)}}\ell_{Pair}(\bm{\theta}|\bm{a},\bm{Z})\\
     &=\sum_{t=1}^{T-1}\sum_{s=t+1}^{T}p_{ts}\log\Bigg[\prod_{j_t=1}^K \prod_{j_s=1}^K \\
     &\bigg( f_{\bm{A}_t,\bm{A}_s}(\bm{a}_t,\bm{a}_s|\bm{\Sigma}_t=\bm{S}_{j_t}, \bm{\Sigma}_s=\bm{S}_{j_s}, \bm{\Phi})\times \omega_{j_t}\omega_{j_s}\bigg)^{T_{\binom{j_t,j_s}{t,s}}^{(r)}}\Bigg]
    \end{aligned}
\end{equation}

\textbf{M-Step:} We obtain $\bm{\theta}^{(r+1)}$ via maximizing $Q_{Pair}(\bm{\theta}|\bm{\theta}^{(r)})$ with respect to $\bm{\theta}$. To be specific, $\bm{\omega}$ and $\bm{\Phi}$ can be parallelly estimated. The estimate of the weight $\bm{\omega}$ has an analytic expression, expressed as
\begin{equation}
    \omega_k^{(r+1)}=\frac{\sum_{t=1}^{T-1}\sum_{s=t+1}^{T}\sum_{j_t=1}^K T_{\binom{j_t,k}{t,s}}^{(r)}}{\sum_{t=1}^{T-1}\sum_{s=t+1}^{T}\sum_{j_t=1}^K \sum_{j_s=1}^KT_{\binom{j_t,j_s}{t,s}}^{(r)}}.
\end{equation}
$\bm{\Phi}^{(r+1)}$ is obtained via maximizing the function
\begin{equation}
\begin{aligned}
 &Q_{Pair}(\bm{\Phi}|\bm{\Phi}^{(r)})=\\
 &\sum_{t=1}^{T-1}\sum_{s=t+1}^{T}p_{ts}\log\Bigg[\prod_{j_t=1}^K \prod_{j_s=1}^K \\
 &\bigg( f_{\bm{A}_t,\bm{A}_s}(\bm{a}_t,\bm{a}_s|\bm{\Sigma}_t=\bm{S}_{j_t}, \bm{\Sigma}_s=\bm{S}_{j_s}, \bm{\Phi})\bigg)^{T_{\binom{j_t,j_s}{t,s}}^{(r)}}\Bigg],
\end{aligned}
\end{equation}
with respect to $\bm{\Phi}$. The function $Q_{Pair}(\bm{\Phi}|\bm{\Phi}^{(r)})$ can be feasibly maximized using the Quasi-Newton method \citep{dennis1977quasi}.

We obtain the maximum likelihood estimate $\hat{\bm{\theta}}$ until convergence. The conditional expectation of $\mathbb{I}(Z_t=k)$, denoted as $G_{t,k}$, can be used as the classifier to show the probability that the image $t$ belongs to the group $k$, expressed as
\begin{equation}
\begin{aligned}
 &G_{t,k}=\mathbb{E}_{\bm{Z}|\bm{A},\hat{\bm{\theta}}}\mathbb{I}(Z_t=k)=\\
 &\frac{\sum_{s=1}^T\sum_{j_s=1}^K\bigg( f(\bm{a}_t,\bm{a}_s|\bm{\Sigma}_t={\bm{S}_{k}}, \bm{\Sigma}_s=\bm{S}_{j_s}, \hat{\bm{\Phi}}) \hat{\omega}_{j_t}{\hat{\omega}_{k}}\bigg)^{p_{ts}}}{\sum_{s=1}^T\sum_{j_t=1}^K \sum_{j_s=1}^K \bigg( f(\bm{a}_t,\bm{a}_s|\bm{\Sigma}_t=\bm{S}_{j_s}, \bm{\Sigma}_s=\bm{S}_{j_s},  \hat{\bm{\Phi}}) \hat{\omega}_{j_t}\hat{\omega}_{j_s}\bigg)^{p_{ts}}}. 
\end{aligned}
\end{equation}
We classify the observation $t$ to the group $k_t=\max_{k=1,2,...,K}G_{t,k}$. 

The composite likelihood-based EM algorithm borrows the ideas from the two popular statistical methods. We are concerned about if our algorithm still enjoys the properties of the original methods. Fortunately, there have been many relevant works \citep{choi2011composite,gao2011composite,vasdekis2012composite,chen2016composite} endorsing our proposed algorithm. The major concern is that whether the hybrid method still enjoys the three key properties of standard EM algorithm. As shown in detail by \citet{gao2011composite}, it is easy to prove the three key properties which are (i) the ascent property that is $\ell_{Pair}(\bm{\theta}^{{(r+1)}}|\bm{a})>\ell_{Pair}(\bm{\theta}^{(r)}|\bm{a})$, where $\ell_{Pair}(\bm{\theta}|\bm{a})=\int\ell_{Pair}(\bm{\theta}^{{(r+1)}}|\bm{a},\bm{Z})d\bm{Z}$, (ii) convergence to a stationary point of the objective function, and (iii) convergence rate depending on the curvature of the objective function. The \code{MATLAB} codes implementing this hybrid algorithm are attached in the Appendix \ref{sec:codes}.

\section{Simulation Studies}
\label{sec:numerical}
In this section, we use synthetic data to study the performance of our proposed algorithm.  The synthetic data are generated based on the full model (Model \ref{eq:model}), where the correlation function is the exponential function, i.e., $\mathcal{K}(d;\phi)=\exp(-d/\phi)$. The range parameter $\phi>0$ controls the correlation, and a larger $\phi$ leads to a larger correlation. In each replication, $T=50$ positive definite matrices are generated for classification. The \textit{remaining} terms $U_t$ are generated by setting the degrees of freedom as $M=5$ and setting range parameter as $\phi=1$. The latent variable $Z_t$ are sampled equally from $\{1,2,3\}$, indicating $K=3$ and $\omega_1=\omega_2=\omega_3=\frac{1}{3}$. In each replication, the mean $\bm{S}_k$ is independently sampled from $\mathcal{W}_3(\bm{I},3)$. The covariate is $\bm{X}_t=[X_{t1}, ..., X_{td}, ..., X_{t,10}]$, and $X_{td}$ is sampled from a uniform distribution ranging from 0 to 1. Subsequently, all the entries in the distance matrix measuring all the pairs of the subjects are also scaled to $[0,1]$. For each comparison, $100$ replications are generated.

To have a fair comparison, we do not use the true means $\{\bm{S}_k:k=1,2,...,K\}$ as the input of our algorithm. Instead, we give the trained mean as followed. For each given $\bm{S}_k$, we generate $10$ positive definite matrices according to our model, and then we calculate their sample mean. The sample means, denoted as $\{\tilde{\bm{S}}_k:k=1,2,...,K\}$, are the algorithm inputs. Using $\{\tilde{\bm{S}}_k:k=1,2,...,K\}$ is also better mimicking the real world.

\citet{huang2015log} uses the Log-Euclidean metric for image set classification. The Log-Euclidean metric has already been used as a metric measuring the distance between positive definite matrices in many other studies \citep[e.g.,][]{arsigny2006log}. In our study, we implement this method in a simplified way, that is to give the group of the image $t$ via finding a $k$ which minimizes the function $||\log \bm{A}_t-\log \tilde{\bm{S}}_k||_F$. Since both our algorithm and the method of Log-Euclidean metric utilize the information from the training data, i.e., $\{\tilde{\bm{S}}_k:k=1,2,...,K\}$, the method of Log-Euclidean metric can be a benchmark method to tell the improvement caused by modeling the correlation caused by the auxiliary information. 

{We also compare our proposal to the other two compelling Wishart distribution-based methods. The first one is the Wishart mixture model relying on Expectation-Maximization algorithm, proposed by \citet{hidot2010expectation}. To make the Wishart mixture model comparable to our proposal, we fixed the mixture means as $\{\tilde{\bm{S}}_k:k=1,2,...,K\}$. In this way, the Wishart mixture model can be considered as a our proposal by setting $\rho_{ts}=0$ for all $t<s$. Thus, it can be used to validate if the additive Wishart process makes improvements. A compelling benchmark method is the Wishart BNP proposed by \citet{cherian2016bayesian}, which is a Dirichlet process of Wishart distribution. The computation of the Wishart BNP is based on Gibbs sampling. Similar to our algorithm, the Wishart BNP was also proposed with the application to the RCD-based image set classification. The Wishart BNP does not require training data and produces classification with an undetermined number of groups. Therefore, we further report the number of groups identified by the algorithm of the Wishart BNP.}

The methods we give above utilize the whole matrix. Alternatively, we may only use the eigenvalues of an RCD. By fixing the number of groups as $K=3$, we can classify the images by the K-means and the Gaussian mixture model, treating the eigenvalues as the multivariate responses. The properties of our proposed method (EM Hybrid) and the benchmark methods are summarized in Table \ref{tab:methods}, helping us learn the key differences among the methods.

\begin{table}[t]
\caption{The proposed method (EM Hybrid) and the benchmark methods are summarized. The properties of these methods are number of groups (fixed v.s. flexible), training data request (yes v.s. no), model based (yes v.s. no), and information used (whole matrix v.s. eigenvalues). }\label{tab:methods}
\centering
\footnotesize
\begin{tabular}{c|cccc}
\hline\hline
Method & \begin{tabular}[c]{@{}c@{}}Number\\ of Groups\end{tabular} & \begin{tabular}[c]{@{}c@{}}Training Data \\ Request\end{tabular} & \begin{tabular}[c]{@{}c@{}}Model \\ Based\end{tabular} & \begin{tabular}[c]{@{}c@{}}Information\\ Used\end{tabular} \\ \hline
EM Hybrid & Fixed & Yes & Yes & Whole \\
{Wishart Mixture} & {Fixed} & {Yes} & {Yes} & {Whole} \\ 
Wishart BNP & Flexible & No & Yes & Whole \\
Log-Euclidean & Fixed & Yes & No & Whole \\
K-means & Fixed & No & No & Eigenvalues \\
Gaussian Mixture & Fixed & No & Yes & Eigenvalues \\ \hline\hline
\end{tabular}
\end{table}

\begin{figure}[t]
    \centering
    \includegraphics[width=0.5\textwidth]{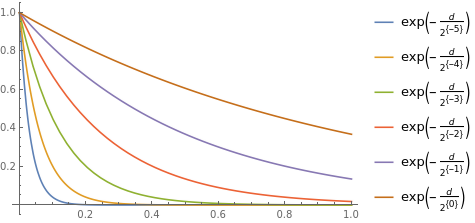}
    \caption{The exponential component in the weight function is displayed.}
    \label{fig:exp}
\end{figure}

We inspect the performance of our algorithm in two aspects: (a) the classification accuracy and (b) the role of the tuning parameters $\lambda$ and $u$ in the likelihood-weight function. To explore the roles of the tuning parameters, we give the combinations of $\lambda\in\{5\times 2^{-5}, 5\times 2^{-4}, 5\times 2^{-3},5\times 2^{-2},5\times 2^{-1},5\times 2^{0}\}$, and $u\in\{0.2,0.4,0.6,0.8\}$. The exponential component in the weight function is displayed in Figure \ref{fig:exp}.

The methods to be compared include the supervised methods (i.e., EM Hybrid and Log-Euclidean Metric), the semi-supervised methods (i.e., Gaussian mixture Model and K-means), and the unsupervised method (i.e., Wishart BNP), To make the classification accuracy measured under the same system, we use the commonly-used Rand index  as metric to describe the classification accuracy \citep{hubert1985comparing}. The Rand index \citep{rand1971objective} is a value ranging from 0 to 1, and a larger value indicates that the classification is more accurate. 

\begin{figure*}[ht!]
    \centering
         \includegraphics[width=1.05\textwidth]{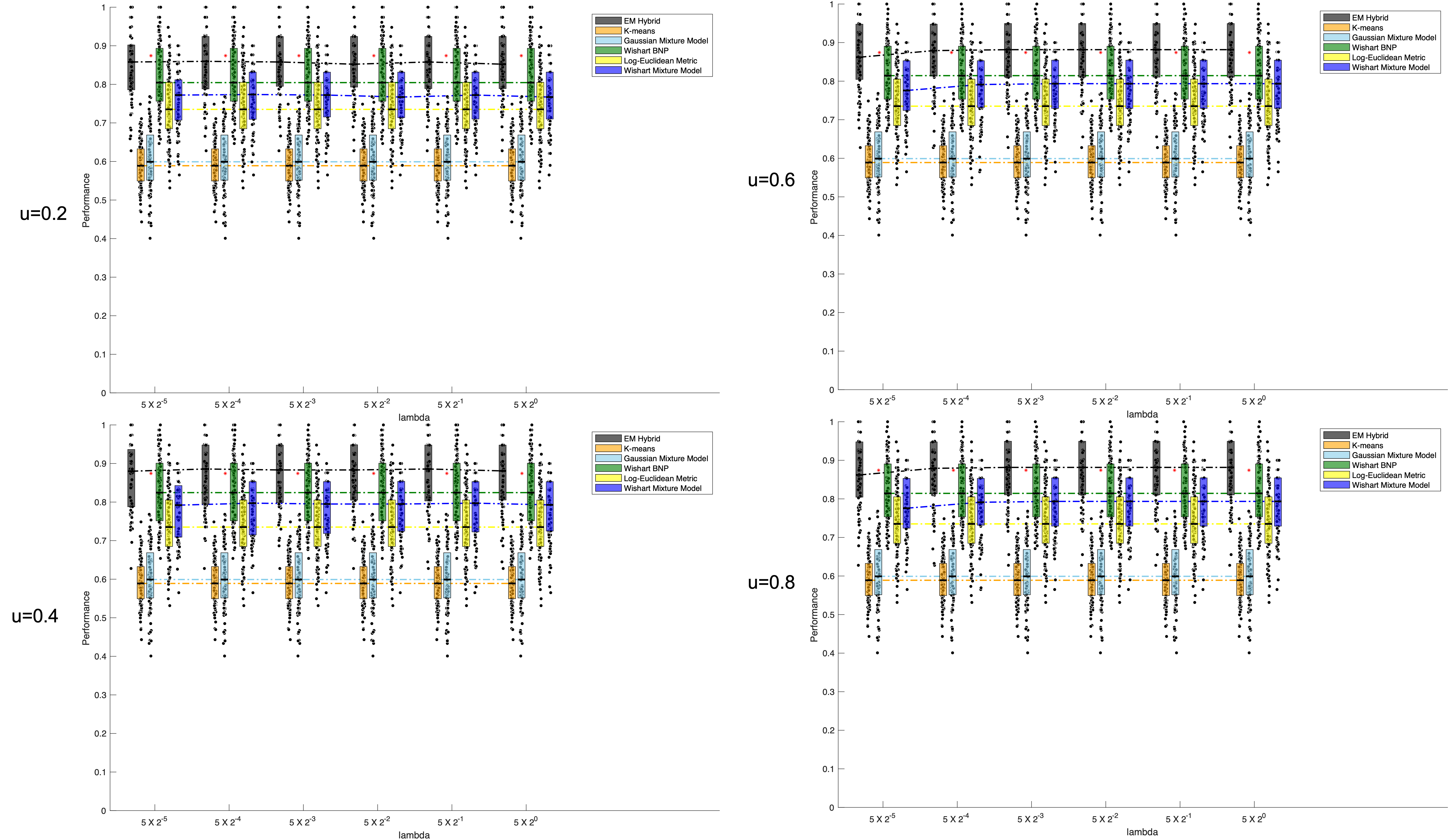}
    \caption{The side-by-side boxplots of the Rand indices are displayed. The title of each subplot gives the value of $u$. Under a certain $u$, the x-axis is for the values of $\lambda$ and the y-axis is for the Rand index. The Rand indices produced by a method of all replications are visualized by a boxplot. For each method, the medians of the boxplots are connected by a dashed line. The legends on the right side give the colors for each method. To clarify, the benchmark methods do not rely on $\lambda$ and $u$ but are only visualized at the locations to compare with the corresponding Rand indices produced by our algorithm.}
    \label{fig:comparison}
\end{figure*}

First, we compare our algorithm to the other benchmark methods, in terms of classification accuracy. The side-by-side boxplots of the Rand indices are displayed in Figure \ref{fig:comparison}. From the results, our algorithm with different settings of tuning parameters produces a more accurate classification result than the others. We also note that the method using only eigenvalues both produce worse results, compared to the ones which utilize the whole matrix. The performance of the Log-Euclidean metric method shows that the inclusion of the variations caused by the auxiliary information is important. As we stated before, the Wishart BNP is a compelling method. In our simulation study, the performance of our algorithm is better than the Wishart BNP. {We collect the posterior mode of the clusters in the Wishart BNP. Over all the replications, the averaged posterior mode is 5.74. Among the 50 replications, only 15 replications have the posterior mode estimated as 3. This indicates that the Wishart BNP can be over the true clusters. The Wishart mixture model does not perform as good as our proposal, indicating the importance of inducing correlation of RCDs.} 

\begin{figure}[ht!]
    \centering
         \includegraphics[width=0.5\textwidth]{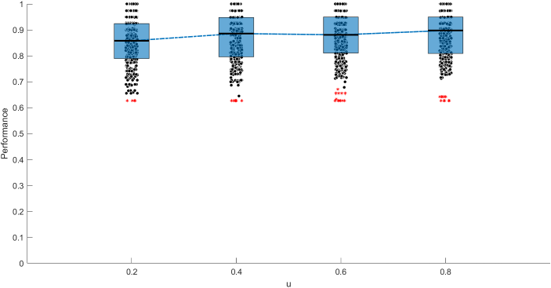}
    \caption{The side-by-side boxplots of the Rand indices are displayed. The x-axis is for the values of $u$ and the y-axis is for the Rand index. Given a $u$, the rand indices are aggregated over all $\lambda\in\{5\times 2^{-5}, 5\times 2^{-4}, 5\times 2^{-3},5\times 2^{-2},5\times 2^{-1},5\times 2^{0}\}$. The aggregated Rand indices are visualized by a boxplot. The medians of the boxplots are connected by a dashed line. }
    \label{fig:u}
\end{figure}

{A nice feature of our algorithm is that our algorithm is robust to the tuning parameters, in terms of classification accuracy. Given the demo script provided in Appendix \ref{sec:codes}, the typical computational time of $u=0.1$ and $T=50$ is 1 minute. However, we still recommend to give a larger $u$ if the large computational cost allows in a certain problem, because we observe that a larger $u$ leads to larger Rand indices in Figure \ref{fig:u} where the Rand indices are aggregated over all $\lambda\in\{5\times 2^{-5}, 5\times 2^{-4}, 5\times 2^{-3},5\times 2^{-2},5\times 2^{-1},5\times 2^{0}\}$. Also, we recommend authors to carry out some pilot studies to choose the optimal $\lambda$. }

\begin{figure*}[ht!]
\centering
         \includegraphics[width=0.7\textwidth]{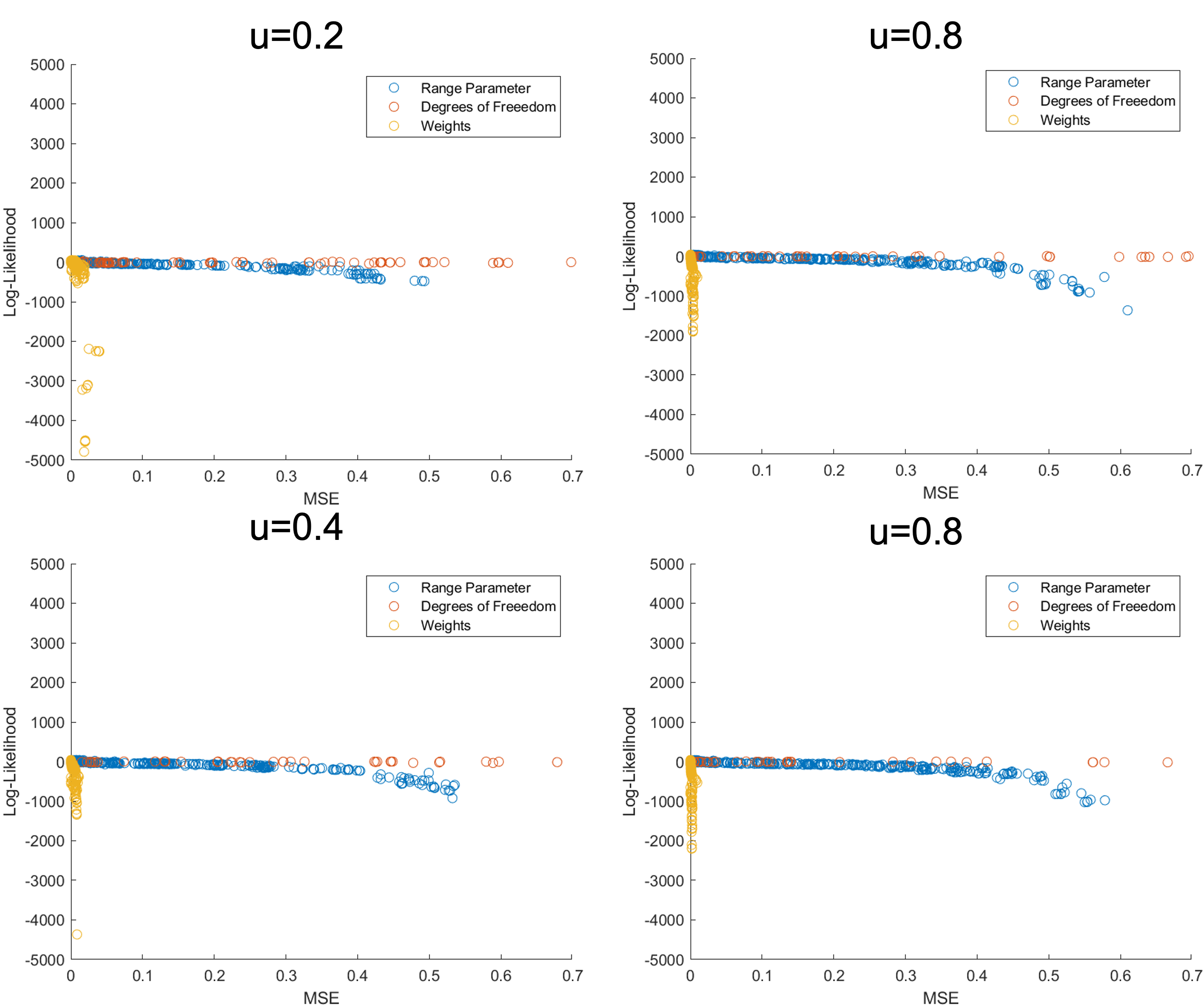}
    \caption{We use scatter plot to show the relationship between the {squared errors} and the log-likelihood. The title of each subplot gives the value $u$. The x-axis is for the values of {squared errors} and the y-axis is for the values of log-likelihood. The legends on the right side give the colors for each parameter.}
    \label{fig:para}
\end{figure*}

Furthermore, we want to investigate the role of the tuning parameters on parameter estimation. The parameter estimations are evaluated by the {squared error}. The {squared error} are defined as follows:
\begin{itemize}
    \item \textbf{Group-Weights:} $\frac{1}{K}\sum_{k=1}^K(\hat{\omega}_k-\omega_k)^2$
    \item \textbf{Range Parameter:} $(\hat{\phi}-\phi)^2$
    \item \textbf{Degrees of Freedom:} $(\hat{M}-M)^2$,
\end{itemize}
where the terms without hats are the true values of the parameters and the terms with hats are the estimates. We use dot plots to show the relationship between the {squared errors} and the log-likelihood (Figure \ref{fig:para}). Generally, we observe a negative association between the {squared errors} and the log-likelihood. In light of the negative association, we recommend reporting the parameter estimation whose log-likelihood value is largest.

\section{Application to the Chicago Face Database} 
\label{sec:application}

We use the Chicago face database \citep{ma2015chicago} as the motivating data-set. The data-set is open to the public. It can be requested from the website \url{https://chicagofaces.org}. The data-set has headshots of the subjects of different races, together with the subjects' auxiliary information (e.g., age, nose width, etc). Our goal is to classify the images by their races (i.e., Black, Latino, and White), through using their RCDs. Race classification is one of the common scientific goals in computer vision studies \citep[e.g.,][]{ou2005real,roomi2011race,fu2014learning}. The true races of the subjects are known in our motivating data-set, thus we can evaluate the classification accuracy by comparing the inferred race classification and the true race classification.

The RCDs are calculated as follows. First we only use the voxels which cover the skin to calculate the RCD. This step can be simply implemented by the \code{MATLAB} codes provided by \citet{Gaurav_2020}. The feature vector is then defined as $\bm{F}(v)=[F(v,R), F(v,G), F(v,B)]^T$, where $F(v,R), F(v,G), F(v,B)$ are the magnitudes of red, green, and blue at the voxel $v$. We further normalized the magnitudes by dividing the image's sample standard error of the magnitudes at the blue channel, denoted as $\tilde{\bm{F}}(v)=\bm{F}(v)/\sigma_B=[F(v,R), F(v,G), F(v,B)]^T/\sigma_B$. Finally, the RCD is computed by taking the sample covariance of all $\tilde{\bm{F}}(v)$. 

Note that the original RCD paper \citep{tuzel2006region} only introduced a big picture, and the scheme to generate the RCD varies under different scenarios. In short, the scheme to generate the RCDs relies on the relevant domain knowledge. Our paper primarily focuses on introducing the proposed methodology. Thus we simply use the scheme described above, and the scheme is considerably useful in our classification problem. However, other appropriate schemes \citep[e.g.,][]{roomi2011race,cherian2016bayesian} to generate the RCDs can also be used, if the features extracted are useful in the specific classification problem. More importantly, the other schemes also have no hurdle to be used in our algorithm.

After obtaining the RCDs of the subjects, we select the training data to obtain the trained means $\{\tilde{\bm{S}}_k:k=1,2,...,K\}$. For each race (i.e., Black, Latino, and White), we randomly select one-fifth of the subjects of each race and calculate their sample mean treated as the trained mean $\{\tilde{\bm{S}}_k:k=1,2,...,K\}$. We use all the available numerical auxiliary information to construct $\bm{X}_t$. Each covariate is normalized, and the entries in the distance matrix measuring all the pairs of the subjects are also scaled to $[0,1]$.

\begin{figure*}[ht!]
    \centering
         \includegraphics[width=1.05\textwidth]{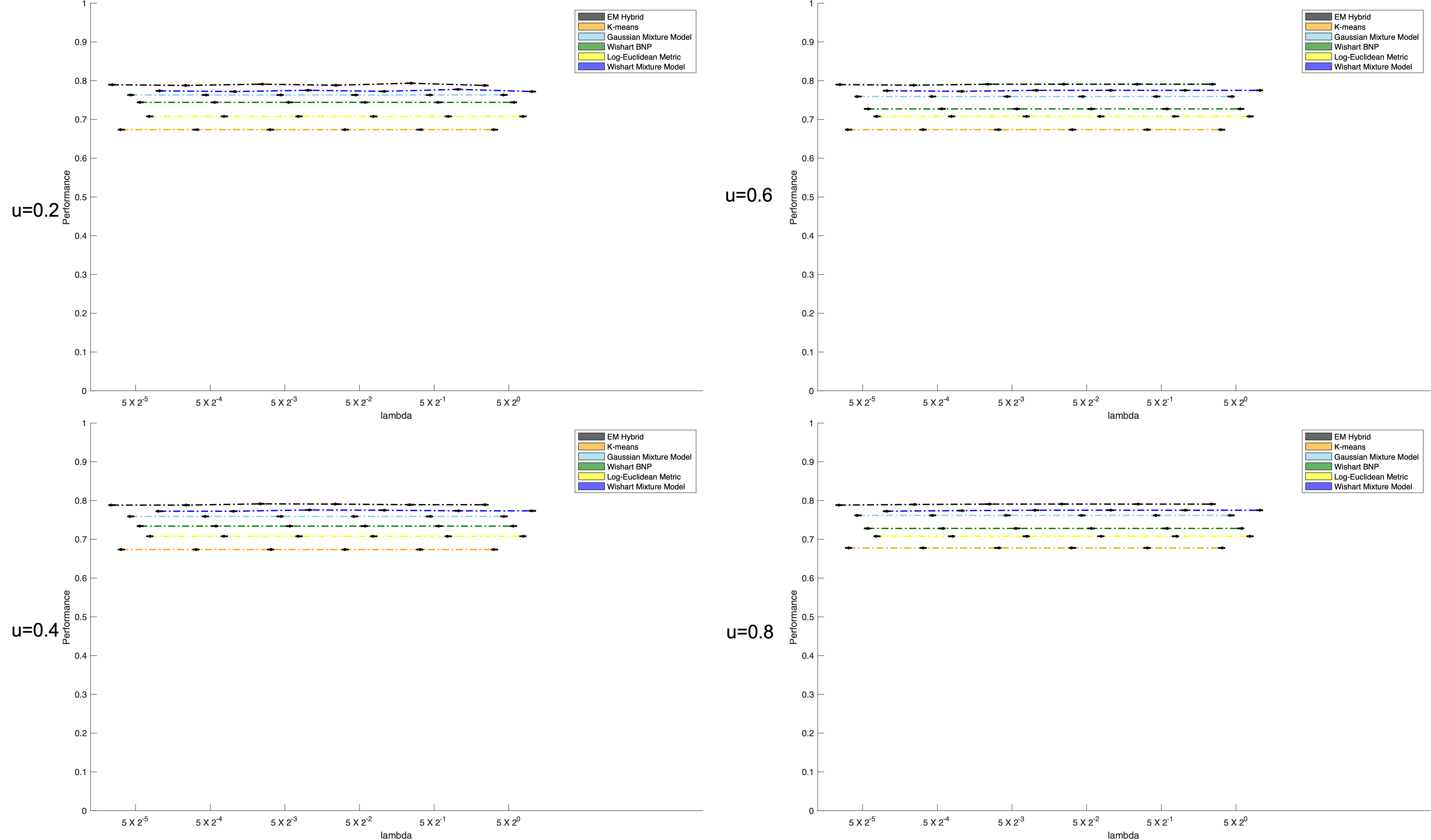}
    \caption{The side-by-side dot-plots of the Rand indices are displayed. The title of each subplot gives the value of $u$. Under a certain $u$, the x-axis is for the values of $\lambda$ and the y-axis is for the Rand index. The Rand index produced by a method is visualized by a dot. For each method, the dots are connected by a dashed line. The legends on the right side give the colors for each method. To clarify, the benchmark methods do not rely on $\lambda$ and $u$ but are only visualized at the locations to compare with the corresponding Rand index produced by our algorithm.}
    \label{fig:real}
\end{figure*}

We continue to use the methods listed in Table \ref{tab:methods}. We also continue to give the combinations of $\lambda\in\{5\times 2^{-5}, 5\times 2^{-4}, 5\times 2^{-3},5\times 2^{-2},5\times 2^{-1},5\times 2^{0}\}$, and $u\in\{0.2,0.4,0.6,0.8\}$. The Rand index is used to measure the classification accuracy too. The classification accuracy in terms of the Rand index is visualized in Figure \ref{fig:real}. The real data result is consistent with what we have had in the simulation study (Section \ref{sec:numerical}). First, our algorithm still holds the first place among all the methods. Also, the classification result is robust to the tuning parameters in the likelihood-weight function. In the real data analysis, the model-based methods (EM Hyrbid, Wishart BNP, {Wishart Mixture Model}, and Gaussian Mixture Model) are better than the non-model-based methods (Log-Euclidean Metric and K-means). This may imply the importance of including statistical uncertainties in modeling real data. The Gaussian mixture model performs relatively well. This may be caused by that the eigenvalues provide distinguishing information in this classification problem.

In light of the suggestion provided in Section \ref{sec:numerical}, the parameter estimation are given as follows. The estimation of the weights are $\hat{\omega}_1=0.3333$, $\hat{\omega}_2=0.3334$, and $\hat{\omega}_3=0.3333$; the parameter estimation for the range parameter is $\hat{\phi}=0.5647$; the parameter estimation for the degrees of freedom is $\hat{M}=39.1730$. The associated tuning parameters are $u=0.4$ and $\lambda=5\times2^{-5}$, whose log-likelihood which is the largest among all settings.


\section{Conclusion and Future Directions}
\label{sec:conclusion}
In this paper, we propose a novel EM composite likelihood-based algorithm of Wishart matrices to classify the Wishart matrices which are correlated caused by the auxiliary information. The methodological details are explicitly given in the paper. The algorithm also performs better than the other methods in terms of image set classification. In summary, our paper has significant contributions in both the methodological development of matrix-variate analysis and the application of computer vision.

The proposed hierarchical model is set up in a relatively simplified way. More complicated features can be added given specific computer vision problems. For example, the mean of the positive definite matrices may also be determined by the subject-level covariates, but not necessarily be plug-in values. However, the added complicated features shall make the statistical inference more challenging. Statistical inference tools such as variational Bayesian methods and Markov chain Monte Carlo methods are encouraged to be implemented in the future.

\section*{Acknowledgement}
{We thank Professor Brian J Reich at North Carolina State University for providing valuable suggestions. We finally thank the Center for Decision Research, The University of Chicago for providing the motivating data, allowing methodological development in computer vision.
}


\appendix

\section{Derivations of the \textit{Bivariate} Density}
\label{sec:derivation}
This scheme to obtain the joint density borrows the idea from \citep{smith2007distribution} in which the joint density for complex Wishart matrices is derived. First, we give the joint density function of $[\bm{U}_t,\bm{U}_s]$ specified in our model (Model \ref{eq:model}), denoted as $f_{\bm{U}_t,\bm{U}_s}[\bm{u}_t,\bm{u}_s]$. The function $f_{\bm{U}_t,\bm{U}_s}(\bm{u}_t,\bm{u}_s)$ can be decomposed as
\begin{equation}
    f_{\bm{U}_t,\bm{U}_s}(\bm{u}_t,\bm{u}_s)=f_{\bm{U}_t|\bm{U}_s}(\bm{u}_t|\bm{u}_s)\times f_{\bm{U}_s}(\bm{u}_s),
\end{equation}
where $f_{\bm{U}_t|\bm{U}_s}(\bm{u}_t|\bm{u}_s)$ is the conditional density function of $[\bm{U}_t|\bm{U}_s]$ and $f_{\bm{U}_s}(\bm{u}_s)$ is the marginal density of $\bm{U}_s$. 

In Section \ref{sec:setup}, $\bm{G}_{jt}|\bm{G}_{js}$ follows a normal distribution such that $\bm{G}_{jt}|\bm{G}_{js}\sim\mathcal{N}(\rho_{ts}\bm{G}_{js},\sqrt{1-\rho_{ts}^2}\bm{I})$. Given \citet[][Equation 67]{james1964distributions}, we can give the explicit expression of $f_{\bm{U}_t|\bm{U}_s}(\bm{u}_t|\bm{u}_s)$\footnote{A similar derivation has been done in \citet{smith2007distribution} for complex correlated Wishart matrices.}, which is expressed as 
\begin{equation}
\small
    \begin{aligned}
    &f_{\bm{U}_t|\bm{U}_s}(\bm{u}_t|\bm{u}_s)=\\
    &\exp\left[-\frac{1}{2}tr\left(\frac{M\rho_{ts}^2}{1-\rho_{ts}^2}\bm{u}_s\right)\right]\times\ _0F_1\left(\frac{1}{2}M;\frac{1}{4}\left(\frac{M\rho_{ts}}{1-\rho_{ts}^2}\right)^2\bm{u}_s\bm{u}_t\right)\\
    &\times\frac{1}{2^{Mp/2}\times\Gamma_p(\frac{1}{2}M)\times|\frac{1-\rho_{ts}^2}{M}\bm{I}|^{M/2}}\times\\
    &\exp\left[-\frac{1}{2}tr\left(\frac{M}{1-\rho_{ts}^2}\bm{u}_t\right)\right]\times |\bm{u}_t|^{\frac{1}{2}(M-p-1)},
    \end{aligned}
\end{equation}
where is $_0F_1(;)$ a hypergeometric function of matrix argument. Then the joint density function $f_{\bm{U}_t,\bm{U}_s}(\bm{u}_t,\bm{u}_s)$ is

\begin{equation}
\footnotesize
    \begin{aligned}
    &f_{\bm{U}_t,\bm{U}_s}(\bm{u}_t,\bm{u}_s)={f_{\bm{U}_t|\bm{U}_s}(\bm{u}_t|\bm{u}_s)}\times {f_{\bm{U}_s}(\bm{u}_s)}\\
    &={\exp\left[-\frac{1}{2}tr\left(\frac{M\rho_{ts}^2}{1-\rho_{ts}^2}\bm{u}_s\right)\right]\times\ _0F_1\left(\frac{1}{2}M;\frac{1}{4}\left(\frac{M\rho_{ts}}{1-\rho_{ts}^2}\right)^2\bm{u}_s\bm{u}_t\right)}\\
    &{\times\frac{1}{2^{Mp/2}\times\Gamma_p(\frac{1}{2}M)\times|\frac{1-\rho_{ts}^2}{M}\bm{I}|^{M/2}}\times}\\
    &{\exp\left[-\frac{1}{2}tr\left(\frac{M}{1-\rho_{ts}^2}\bm{u}_t\right)\right]\times |\bm{u}_t|^{\frac{1}{2}(M-p-1)}}\\
    &{\times |\bm{u}_s|^{\frac{1}{2}(M-p-1)}\times \exp\left(-\frac{1}{2}tr\left(\bm{u}_sM\right)\right)}\\
    &{\times\left(2^{Mp/2}\times|\bm{I}/M|^{M/2}\times\Gamma_{p}\left(\frac{1}{2}M\right)\right)^{-1}}\\
    &=\ _0F_1\left(\frac{1}{2}M;\frac{1}{4}\left(\frac{M\rho_{ts}}{1-\rho_{ts}^2}\right)^2\bm{u}_s\bm{u}_t\right)\times 2^{-Mp} \times \left[\Gamma_p(\frac{1}{2}M)\right]^{-2}\\
    &\times |\bm{u}_t|^{\frac{1}{2}(M-p-1)}\times |\bm{u}_s|^{\frac{1}{2}(M-p-1)} \\
    &\times\left(\frac{1-\rho_{ts}^2}{M^2}\right)^{-pM/2} \times \exp\left(-\frac{1}{2}tr\left[M\frac{1}{1-\rho_{ts}^2}(\bm{u}_s+\bm{u}_t)\right]\right).
    \end{aligned}
\end{equation}

Next, we want to derive the expression of the joint density function $f_{\bm{A}_t,\bm{A}_s}(\bm{a}_t,\bm{a}_s|\bm{\Sigma}_t=\bm{S}_{j_t},\bm{\Sigma}_s=\bm{S}_{j_s},\bm{\Phi})$. This can be done via variable change. We have the relationship that $\bm{U}_t=\bm{Q}_{j_t}^{-1}\bm{A}_t(\bm{Q}_{j_t}^T)^{-1}$. {The Jacobian is $|\bm{J}|=|\bm{\Sigma}_t^{-1}|^{(p+1)/2}\times |\bm{\Sigma}_s^{-1}|^{(p+1)/2}$, given Proposition 5.11 of \citet[][Chapter 5: Matrix Factorizations and Jacobians]{eaton1983multivariate}.} Therefore, the expression of $f_{\bm{A}_t,\bm{A}_s}(\bm{a}_t,\bm{a}_s|\bm{\Sigma}_t=\bm{S}_{j_t},\bm{\Sigma}_s=\bm{S}_{j_s},\bm{\Phi})$ is
\begin{equation}
\footnotesize
    \begin{aligned}
&f_{\bm{A}_t,\bm{A}_s}(\bm{a}_t,\bm{a}_s|\bm{\Sigma}_t=\bm{S}_{j_t},\bm{\Sigma}_s=\bm{S}_{j_s},\bm{\Phi})\\
    &=\ _0F_1\left(\frac{1}{2}M;\frac{1}{4}\left(\frac{M\rho_{ts}}{1-\rho_{ts}^2}\right)^2\bm{Q}_{j_s}^{-1}\bm{a}_s(\bm{Q}_{j_s}^T)^{-1}\bm{Q}_{j_t}^{-1}\bm{a}_t(\bm{Q}_{j_t}^T)^{-1}\right)\\
    &\times 2^{-Mp} \times \left[\Gamma_p(\frac{1}{2}M)\right]^{-2}\times |\bm{S}_{j_t}^{-1}\bm{a}_t|^{\frac{1}{2}(M-p-1)}\times |\bm{S}_{j_s}^{-1}\bm{a}_s|^{\frac{1}{2}(M-p-1)} \\
    &\times\left(\frac{1-\rho_{ts}^2}{M^2}\right)^{-\frac{pM}{2}} \times \exp\left(-\frac{1}{2}tr\left[M\frac{1}{1-\rho_{ts}^2}\left(\bm{S}_{j_s}^{-1}\bm{a}_s+\bm{S}_{j_t}^{-1}\bm{a}_t\right)\right]\right)\\
    &\times |\bm{S}_{j_t}^{-1}|^{\frac{p+1}{2}}\times |\bm{S}_{j_s}^{-1}|^{\frac{p+1}{2}},
    \end{aligned}
\end{equation}

\section{Codes}
\label{sec:codes}
The \code{MATLAB} codes are in \code{EM$\_$Hybrid$\_$Codes.zip}. The instructions of implementing the codes are provided \code{readme.txt}. Some example scripts are also given.


\end{document}